\def\Version{ 1.02
  }







\message{Assuming 8.5" x 11" paper and .eps figures}    

\magnification=\magstep1	          

\raggedbottom

\parskip=9pt

%

\def\singlespace{\baselineskip=12pt}      
\def\sesquispace{\baselineskip=16pt}      


 







\font\openface=msbm10 at10pt
 %

\def\Reals         {{\hbox{\openface R}}}

 %
 %
 %




\font\german=eufm10 at 10pt

\def\Buchstabe#1{{\hbox{\german #1}}}




\def\Re  {\mathop{\rm Re}  \nolimits}    
\def\Imag  {\mathop{\rm Imag}  \nolimits}



\def\tr{\mathop {{\rm \, Tr}} \nolimits}	 


%

%

%
%



\def\implies{\Rightarrow}

%




\def\sqr#1#2{\vcenter{
  \hrule height.#2pt 
  \hbox{\vrule width.#2pt height#1pt 
        \kern#1pt 
        \vrule width.#2pt}
  \hrule height.#2pt}}


 %



\def\lto{\mathop
        {\hbox{${\lower3.8pt\hbox{$<$}}\atop{\raise0.2pt\hbox{$\sim$}}$}}}
\def\gto{\mathop
        {\hbox{${\lower3.8pt\hbox{$>$}}\atop{\raise0.2pt\hbox{$\sim$}}$}}}
%
%
%


\def\half{{1 \over 2}}


\def\part{\subseteq}		


\def\braces#1{ \{ #1 \} }



\def\to{\mathop\rightarrow}	

\def\ideq{\equiv}		



\def\interior #1 {  \buildrel\circ\over  #1}     




\def\basisvector#1#2#3{
 \lower6pt\hbox{
  ${\buildrel{\displaystyle #1}\over{\scriptscriptstyle(#2)}}$}^#3}

\def\alfa{\alpha}


\def\hat{\widehat}		







\fontdimen16\textfont2=2.5pt
\fontdimen17\textfont2=2.5pt
\fontdimen14\textfont2=4.5pt
\fontdimen13\textfont2=4.5pt 

\let\miguu=\footnote
\def\footnote#1#2{{$\,$\parindent=9pt\baselineskip=13pt%
\miguu{#1}{#2\vskip -7truept}}}
 %

\def\linebreak{\hfil\break}
\def\lbr{\linebreak}
\def\pagebreak{\vfil\break}


\def\BulletItem #1 {\item{$\bullet$}{#1 }}
\def\bulletitem #1 {\BulletItem{#1}}

\def\REMARK{\noindent {\csmc Remark \ }}

\def\PrintVersionNumber{
 \vskip -1 true in \medskip 
 \rightline{version \Version} 
 \vskip 0.3 true in \bigskip \bigskip}

\def\author#1 {\medskip\centerline{\it #1}\bigskip}

\def\address#1{\centerline{\it #1}\smallskip}

\def\furtheraddress#1{\centerline{\it and}\smallskip\centerline{\it #1}\smallskip}

\def\email#1{\smallskip\centerline{\it address for email: #1}} 

\def\AbstractBegins
{
 \singlespace                                        
 \bigskip\leftskip=1.5truecm\rightskip=1.5truecm     
 \centerline{\bf Abstract}
 \smallskip
 \noindent	
 } 
\def\AbstractEnds
{
 \bigskip\leftskip=0truecm\rightskip=0truecm       
 }

\def\section #1 {\bigskip\noindent{\headingfont #1 }\par\nobreak\smallskip\noindent}

\def\subsection #1 {\medskip\noindent{\subheadfont #1 }\par\nobreak\smallskip\noindent}
 %

\def\ReferencesBegin
{
 \singlespace					   
 \vskip 0.5truein
 \centerline           {\bf References}
 \par\nobreak
 \medskip
 \noindent
 \parindent=2pt
 \parskip=6pt			
 }
 %


\def\reference{\hangindent=1pc\hangafter=1} 

\def\ref{\reference}

 %

\def\journaldata#1#2#3#4{{\it #1\/}\phantom{--}{\bf #2$\,$:} $\!$#3 (#4)}
 %

\def\eprint#1{{\tt #1}}

\def\arxiv#1{\hbox{\tt http://arXiv.org/abs/#1}}
 %


\def\webhome{{\tt http://www.pitp.ca/personal/rsorkin/}}
 %

 %



\def\webtilde{\lower2pt\hbox{${\widetilde{\phantom{m}}}$}}

 %

\def\hpf#1{\webhome{\tt{some.papers/}}}
 %

\def\hpfll#1{\webhome{\tt{lisp.library/}}}
 %



\font\titlefont=cmb10 scaled\magstep2 

\font\headingfont=cmb10 at 12pt
%

\font\subheadfont=cmssi10 scaled\magstep1 
%


\font\csmc=cmcsc10  






\def\bra{\langle}
\def\ket{\rangle}
\def\fihat{{\hat\phi}}
\def\Hilb{{\Buchstabe{H}}}
\def\xptn#1{{\langle #1 \, \rangle}}

\def\A{\Buchstabe{A}}







\edef\resetatcatcode{\catcode`\noexpand\@\the\catcode`\@\relax}
\ifx\miniltx\undefined\else\endinput\fi
\let\miniltx\box

\def\makeatletter{\catcode`\@11\relax}

\makeatletter

\def\@makeother#1{\catcode`#1=12\relax}

\def\@ifnextchar#1#2#3{%
  \let\reserved@d=#1%
  \def\reserved@a{#2}\def\reserved@b{#3}%
  \futurelet\@let@token\@ifnch}
\def\@ifnch{%
  \ifx\@let@token\@sptoken
    \let\reserved@c\@xifnch
  \else
    \ifx\@let@token\reserved@d
      \let\reserved@c\reserved@a
    \else
      \let\reserved@c\reserved@b
    \fi
  \fi
  \reserved@c}
\begingroup
\def\:{\global\let\@sptoken= } \:  
\def\:{\@xifnch} \expandafter\gdef\: {\futurelet\@let@token\@ifnch}
\endgroup

\def\@ifstar#1{\@ifnextchar *{\@firstoftwo{#1}}}
\long\def\@dblarg#1{\@ifnextchar[{#1}{\@xdblarg{#1}}}
\long\def\@xdblarg#1#2{#1[{#2}]{#2}}

\long\def \@gobble #1{}
\long\def \@gobbletwo #1#2{}
\long\def \@gobblefour #1#2#3#4{}
\long\def\@firstofone#1{#1}
\long\def\@firstoftwo#1#2{#1}
\long\def\@secondoftwo#1#2{#2}

\def\NeedsTeXFormat#1{\@ifnextchar[\@needsf@rmat\relax}
\def\@needsf@rmat[#1]{}
\def\ProvidesPackage#1{\@ifnextchar[%
    {\@pr@videpackage{#1}}{\@pr@videpackage#1[]}}
\def\@pr@videpackage#1[#2]{\wlog{#1: #2}}

\let\DeclareOption\@gobbletwo
\def\ProcessOptions{\@ifstar\relax\relax}

\def\RequirePackage{%
  \@fileswithoptions\@pkgextension}
\def\@fileswithoptions#1{%
  \@ifnextchar[
    {\@fileswith@ptions#1}%
    {\@fileswith@ptions#1[]}}
\def\@fileswith@ptions#1[#2]#3{%
  \@ifnextchar[
  {\@fileswith@pti@ns#1[#2]#3}%
  {\@fileswith@pti@ns#1[#2]#3[]}}

\def\@fileswith@pti@ns#1[#2]#3[#4]{%
    \def\reserved@b##1,{%
      \ifx\@nil##1\relax\else
        \ifx\relax##1\relax\else
         \noexpand\@onefilewithoptions##1[#2][#4]\noexpand\@pkgextension
        \fi
        \expandafter\reserved@b
      \fi}%
      \edef\reserved@a{\zap@space#3 \@empty}%
      \edef\reserved@a{\expandafter\reserved@b\reserved@a,\@nil,}%
  \reserved@a}

\def\zap@space#1 #2{%
  #1%
  \ifx#2\@empty\else\expandafter\zap@space\fi
  #2}

\let\@empty\empty
\def\@pkgextension{sty}

\def\@onefilewithoptions#1[#2][#3]#4{%
  \input #1.#4 }

\def\typein{%
  \let\@typein\relax
  \@testopt\@xtypein\@typein}
\def\@xtypein[#1]#2{%
  \message{#2}%
  \advance\endlinechar\@M
  \read\@inputcheck to#1%
  \advance\endlinechar-\@M
  \@typein}
\def\@namedef#1{\expandafter\def\csname #1\endcsname}
\def\@nameuse#1{\csname #1\endcsname}
\def\@cons#1#2{\begingroup\let\@elt\relax\xdef#1{#1\@elt #2}\endgroup}
\def\@car#1#2\@nil{#1}
\def\@cdr#1#2\@nil{#2}
\def\@carcube#1#2#3#4\@nil{#1#2#3}
\def\@preamblecmds{}

\def\@star@or@long#1{%
  \@ifstar
   {\let\l@ngrel@x\relax#1}%
   {\let\l@ngrel@x\long#1}}

\let\l@ngrel@x\relax
\def\newcommand{\@star@or@long\new@command}
\def\new@command#1{%
  \@testopt{\@newcommand#1}0}
\def\@newcommand#1[#2]{%
  \@ifnextchar [{\@xargdef#1[#2]}%
                {\@argdef#1[#2]}}
\long\def\@argdef#1[#2]#3{%
   \@ifdefinable #1{\@yargdef#1\@ne{#2}{#3}}}
\long\def\@xargdef#1[#2][#3]#4{%
  \@ifdefinable#1{%
     \expandafter\def\expandafter#1\expandafter{%
          \expandafter
          \@protected@testopt
          \expandafter
          #1%
          \csname\string#1\expandafter\endcsname
          {#3}}%
       \expandafter\@yargdef
          \csname\string#1\endcsname
           \tw@
           {#2}%
           {#4}}}
\def\@testopt#1#2{%
  \@ifnextchar[{#1}{#1[#2]}}
\def\@protected@testopt#1{
  \ifx\protect\@typeset@protect
    \expandafter\@testopt
  \else
    \@x@protect#1%
  \fi}
\long\def\@yargdef#1#2#3{%
  \@tempcnta#3\relax
  \advance \@tempcnta \@ne
  \let\@hash@\relax
  \edef\reserved@a{\ifx#2\tw@ [\@hash@1]\fi}%
  \@tempcntb #2%
  \@whilenum\@tempcntb <\@tempcnta
     \do{%
         \edef\reserved@a{\reserved@a\@hash@\the\@tempcntb}%
         \advance\@tempcntb \@ne}%
  \let\@hash@##%
  \l@ngrel@x\expandafter\def\expandafter#1\reserved@a}
\long\def\@reargdef#1[#2]#3{%
  \@yargdef#1\@ne{#2}{#3}}
\def\renewcommand{\@star@or@long\renew@command}
\def\renew@command#1{%
  {\escapechar\m@ne\xdef\@gtempa{{\string#1}}}%
  \expandafter\@ifundefined\@gtempa
     {\@latex@error{\string#1 undefined}\@ehc}%
     {}%
  \let\@ifdefinable\@rc@ifdefinable
  \new@command#1}
\long\def\@ifdefinable #1#2{%
      \edef\reserved@a{\expandafter\@gobble\string #1}%
     \@ifundefined\reserved@a
         {\edef\reserved@b{\expandafter\@carcube \reserved@a xxx\@nil}%
          \ifx \reserved@b\@qend \@notdefinable\else
            \ifx \reserved@a\@qrelax \@notdefinable\else
              #2%
            \fi
          \fi}%
         \@notdefinable}
\let\@@ifdefinable\@ifdefinable
\long\def\@rc@ifdefinable#1#2{%
  \let\@ifdefinable\@@ifdefinable
  #2}
\def\newenvironment{\@star@or@long\new@environment}
\def\new@environment#1{%
  \@testopt{\@newenva#1}0}
\def\@newenva#1[#2]{%
   \@ifnextchar [{\@newenvb#1[#2]}{\@newenv{#1}{[#2]}}}
\def\@newenvb#1[#2][#3]{\@newenv{#1}{[#2][#3]}}
\def\renewenvironment{\@star@or@long\renew@environment}
\def\renew@environment#1{%
  \@ifundefined{#1}%
     {\@latex@error{Environment #1 undefined}\@ehc
     }{}%
  \expandafter\let\csname#1\endcsname\relax
  \expandafter\let\csname end#1\endcsname\relax
  \new@environment{#1}}
\long\def\@newenv#1#2#3#4{%
  \@ifundefined{#1}%
    {\expandafter\let\csname#1\expandafter\endcsname
                         \csname end#1\endcsname}%
    \relax
  \expandafter\new@command
     \csname #1\endcsname#2{#3}%
     \l@ngrel@x\expandafter\def\csname end#1\endcsname{#4}}

\def\providecommand{\@star@or@long\provide@command}
\def\provide@command#1{%
  {\escapechar\m@ne\xdef\@gtempa{{\string#1}}}%
  \expandafter\@ifundefined\@gtempa
    {\def\reserved@a{\new@command#1}}%
    {\def\reserved@a{\renew@command\reserved@a}}%
   \reserved@a}%

\def\@ifundefined#1{%
  \expandafter\ifx\csname#1\endcsname\relax
    \expandafter\@firstoftwo
  \else
    \expandafter\@secondoftwo
  \fi}

\chardef\@xxxii=32
\mathchardef\@Mi=10001
\mathchardef\@Mii=10002
\mathchardef\@Miii=10003
\mathchardef\@Miv=10004

\newcount\@tempcnta
\newcount\@tempcntb
\newif\if@tempswa\@tempswatrue
\newdimen\@tempdima
\newdimen\@tempdimb
\newdimen\@tempdimc
\newbox\@tempboxa
\newskip\@tempskipa
\newskip\@tempskipb
\newtoks\@temptokena

\long\def\@whilenum#1\do #2{\ifnum #1\relax #2\relax\@iwhilenum{#1\relax
     #2\relax}\fi}
\long\def\@iwhilenum#1{\ifnum #1\expandafter\@iwhilenum
         \else\expandafter\@gobble\fi{#1}}
\long\def\@whiledim#1\do #2{\ifdim #1\relax#2\@iwhiledim{#1\relax#2}\fi}
\long\def\@iwhiledim#1{\ifdim #1\expandafter\@iwhiledim
        \else\expandafter\@gobble\fi{#1}}
\long\def\@whilesw#1\fi#2{#1#2\@iwhilesw{#1#2}\fi\fi}
\long\def\@iwhilesw#1\fi{#1\expandafter\@iwhilesw
         \else\@gobbletwo\fi{#1}\fi}
\def\@nnil{\@nil}
\def\@empty{}
\def\@fornoop#1\@@#2#3{}
\long\def\@for#1:=#2\do#3{%
  \expandafter\def\expandafter\@fortmp\expandafter{#2}%
  \ifx\@fortmp\@empty \else
    \expandafter\@forloop#2,\@nil,\@nil\@@#1{#3}\fi}
\long\def\@forloop#1,#2,#3\@@#4#5{\def#4{#1}\ifx #4\@nnil \else
       #5\def#4{#2}\ifx #4\@nnil \else#5\@iforloop #3\@@#4{#5}\fi\fi}
\long\def\@iforloop#1,#2\@@#3#4{\def#3{#1}\ifx #3\@nnil
       \expandafter\@fornoop \else
      #4\relax\expandafter\@iforloop\fi#2\@@#3{#4}}
\def\@tfor#1:={\@tf@r#1 }
\long\def\@tf@r#1#2\do#3{\def\@fortmp{#2}\ifx\@fortmp\space\else
    \@tforloop#2\@nil\@nil\@@#1{#3}\fi}
\long\def\@tforloop#1#2\@@#3#4{\def#3{#1}\ifx #3\@nnil
       \expandafter\@fornoop \else
      #4\relax\expandafter\@tforloop\fi#2\@@#3{#4}}
\long\def\@break@tfor#1\@@#2#3{\fi\fi}
\def\@removeelement#1#2#3{%
  \def\reserved@a##1,#1,##2\reserved@a{##1,##2\reserved@b}%
  \def\reserved@b##1,\reserved@b##2\reserved@b{%
    \ifx,##1\@empty\else##1\fi}%
  \edef#3{%
    \expandafter\reserved@b\reserved@a,#2,\reserved@b,#1,\reserved@a}}

\let\ExecuteOptions\@gobble

\def\@latex@error#1#2{%
  \errhelp{#2}\errmessage{#1}}

\bgroup\uccode`\!`\%\uppercase{\egroup
\def\@percentchar{!}}

\ifx\@@input\@undefined
 \let\@@input\input
\fi

\def\input{\@ifnextchar\bgroup\@iinput\@@input}
\def\@iinput#1{\input{#1} }

\ifx\filename@parse\@undefined
  \def\reserved@a{./}\ifx\@currdir\reserved@a
    \wlog{^^JDefining UNIX/DOS style filename parser.^^J}
    \def\filename@parse#1{%
      \let\filename@area\@empty
      \expandafter\filename@path#1/\\}
    \def\filename@path#1/#2\\{%
      \ifx\\#2\\%
         \def\reserved@a{\filename@simple#1.\\}%
      \else
         \edef\filename@area{\filename@area#1/}%
         \def\reserved@a{\filename@path#2\\}%
      \fi
      \reserved@a}
  \else\def\reserved@a{[]}\ifx\@currdir\reserved@a
    \wlog{^^JDefining VMS style filename parser.^^J}
    \def\filename@parse#1{%
      \let\filename@area\@empty
      \expandafter\filename@path#1]\\}
    \def\filename@path#1]#2\\{%
      \ifx\\#2\\%
         \def\reserved@a{\filename@simple#1.\\}%
      \else
         \edef\filename@area{\filename@area#1]}%
         \def\reserved@a{\filename@path#2\\}%
      \fi
      \reserved@a}
  \else\def\reserved@a{:}\ifx\@currdir\reserved@a
    \wlog{^^JDefining Mac style filename parser.^^J}
    \def\filename@parse#1{%
      \let\filename@area\@empty
      \expandafter\filename@path#1:\\}
    \def\filename@path#1:#2\\{%
      \ifx\\#2\\%
         \def\reserved@a{\filename@simple#1.\\}%
      \else
         \edef\filename@area{\filename@area#1:}%
         \def\reserved@a{\filename@path#2\\}%
      \fi
      \reserved@a}
  \else
    \wlog{^^JDefining generic filename parser.^^J}
    \def\filename@parse#1{%
      \let\filename@area\@empty
      \expandafter\filename@simple#1.\\}
  \fi\fi\fi
  \def\filename@simple#1.#2\\{%
    \ifx\\#2\\%
       \let\filename@ext\relax
    \else
       \edef\filename@ext{\filename@dot#2\\}%
    \fi
    \edef\filename@base{#1}}
  \def\filename@dot#1.\\{#1}
\else
  \wlog{^^J^^J%
    \noexpand\filename@parse was defined in texsys.cfg:^^J%
    \expandafter\strip@prefix\meaning\filename@parse.^^J%
    }
\fi

\long\def \IfFileExists#1#2#3{%
  \openin\@inputcheck#1 %
  \ifeof\@inputcheck
    \ifx\input@path\@undefined
      \def\reserved@a{#3}%
    \else
      \def\reserved@a{\@iffileonpath{#1}{#2}{#3}}%
    \fi
  \else
    \closein\@inputcheck
    \edef\@filef@und{#1 }%
    \def\reserved@a{#2}%
  \fi
  \reserved@a}
\long\def\@iffileonpath#1{%
  \let\reserved@a\@secondoftwo
  \expandafter\@tfor\expandafter\reserved@b\expandafter
             :\expandafter=\input@path\do{%
    \openin\@inputcheck\reserved@b#1 %
    \ifeof\@inputcheck\else
      \edef\@filef@und{\reserved@b#1 }%
      \let\reserved@a\@firstoftwo%
      \closein\@inputcheck
      \@break@tfor
    \fi}%
  \reserved@a}
\long\def \InputIfFileExists#1#2{%
  \IfFileExists{#1}%
    {#2\@addtofilelist{#1}\@@input \@filef@und}}

\chardef\@inputcheck0

\let\@addtofilelist \@gobble

\def\@defaultunits{\afterassignment\remove@to@nnil}
\def\remove@to@nnil#1\@nnil{}

\newdimen\leftmarginv
\newdimen\leftmarginvi

\newdimen\@ovxx
\newdimen\@ovyy
\newdimen\@ovdx
\newdimen\@ovdy
\newdimen\@ovro
\newdimen\@ovri
\newdimen\@xdim
\newdimen\@ydim
\newdimen\@linelen
\newdimen\@dashdim

\long\def\mbox#1{\leavevmode\hbox{#1}}

\let\@onlypreamble\@gobble

\let\protect\relax

\newdimen\fboxsep
\newdimen\fboxrule

\fboxsep = 3pt
\fboxrule = .4pt

\def\@height{height} \def\@depth{depth} \def\@width{width}
\def\@minus{minus}
\def\@plus{plus}
\def\hb@xt@{\hbox to}

\long\def\@begin@tempboxa#1#2{%
   \begingroup
     \setbox\@tempboxa#1{\color@begingroup#2\color@endgroup}%
     \def\width{\wd\@tempboxa}%
     \def\height{\ht\@tempboxa}%
     \def\depth{\dp\@tempboxa}%
     \let\totalheight\@ovri
     \totalheight\height
     \advance\totalheight\depth}
\let\@end@tempboxa\endgroup

\let\set@color\relax
\let\color@begingroup\relax
\let\color@endgroup\relax
\let\color@setgroup\relax

\let\color@hbox\relax
\let\color@vbox\relax
\let\color@endbox\relax


\begingroup
  \catcode`P=12
  \catcode`T=12
  \lowercase{
    \def\x{\def\rem@pt##1.##2PT{##1\ifnum##2>\z@.##2\fi}}}
  \expandafter\endgroup\x
\def\strip@pt{\expandafter\rem@pt\the}


\def\@input#1{%
  \IfFileExists{#1}{\@@input\@filef@und}{\message{No file #1.}}}

\def\@warning{\immediate\write16}


\def\Gin@driver{dvips.def}
\input graphicx.sty
\resetatcatcode


\resetatcatcode                 


\def\Caption#1{\vbox{

 \leftskip=1.5truecm\rightskip=1.5truecm     
 \singlespace                                
 \noindent #1
 \vskip .25in\leftskip=0truecm\rightskip=0truecm}
 \sesquispace}
 %



\phantom{}


\PrintVersionNumber      


\sesquispace
\centerline{{\titlefont Expressing entropy globally in terms of (4D) field-correlations}\footnote{$^{^{\displaystyle\star}}$}%
%
{To appear in the Proceedings of the Seventh International Conference on Gravitation and Cosmology [ICGC],
 held December 2011 in Goa, India, Journal of Physics Conference Series.
}}

\bigskip


\singlespace			        

\author{Rafael D. Sorkin}
\address
 {Perimeter Institute, 31 Caroline Street North, Waterloo ON, N2L 2Y5 Canada}
\furtheraddress
 {Department of Physics, Syracuse University, Syracuse, NY 13244-1130, U.S.A.}
\email{rsorkin@perimeterinstitute.ca}

\AbstractBegins                              
We express the entropy of a scalar field $\phi$ 
directly in terms of its spacetime correlation function
$W(x,y)=\bra\phi(x)\phi(y)\ket$,
assuming that the higher correlators are of ``Gaussian'' form.
The resulting formula associates an
entropy $S(R)$ to any spacetime region $R\,$; and when $R$ is globally
hyperbolic 
with Cauchy surface $\Sigma\,$, 
$S(R)$ can be interpreted as
the entropy of 
the reduced density-matrix belonging to $\Sigma$.
One acquires in particular a new expression for the entropy of
entanglement across an event-horizon.  
Thanks to its spacetime character, this expression makes sense in a causal
set as well as in a continuum spacetime.
%
%
\AbstractEnds                                

\bigskip



\sesquispace
\vskip -10pt

\section{} 
As usually conceived of, entropy is local in time, being defined
relative to a hypersurface $\Sigma\,$.  To compute such an entropy (at
least in its ``Gibbsian'' guise) one must be able to identify a
density-matrix $\rho(\Sigma)$ which one can plug into the formula
$S=\tr\rho\log\rho^{-1}$.  But according to our current understanding,
quantum fields are in general too singular to admit of meaningful
restriction to lower dimensional subsets of spacetime.  If it is
therefore doubtful whether a concept like 
``state on a hypersurface''
can be well
defined, it is virtually certain that this concept will break down in
the discrete context of a causal set.  For reasons such as these it
would be desirable to define entropy in a more global manner by
associating it with a spacetime region rather than a submanifold of
codimension one.  A more global notion of entropy would also seem
fitting in connection with black holes, whose very definition is global
in character.

Of course, black hole entropy will not be understood fully except
against the background of a theory of quantum gravity and quantum
spacetime.  Nevertheless it seems fair to say that quantum correlations
between the hole's interior and its exterior must contribute to its
entropy.  Perhaps, when suitably understood, some such entropy of
entanglement will even turn out to tell the whole story.  Be that as it
may, a portion of the entanglement entropy can be identified without
invoking full quantum gravity, namely that portion belonging to whatever
``matter'' fields are present in the neighborhood of the horizon.

When we compute such an entropy in the continuum, we obtain an infinite
answer, as is well known.  Introducing a cutoff $\ell$ on the other
hand, we obtain $S_{entanglement}\sim A/\ell^2$, which compares
favorably with the exact formula $S_{BH}=2\pi A/\ell_{p}^2$~, where
$\ell_{p}=\sqrt{8\pi G}$ is the rationalized Planck length.
[1] We
recognize here the familiar area law, but with a coefficient that is not
easily determined because it is not ``universal''.  

On one hand
this
sensitivity of $S$ to the details of the cutoff is to be welcomed
because it lets us learn something about the magnitude of $\ell$ and the
nature of Planck-scale physics.
(Although they are sometimes viewed as uninteresting, non-universal
 quantities are actually the most interesting if we are after the
 microscopic physics!)

But 
on the other hand,
the very fact that the answer is not universal also creates a
difficulty.  When we try to compute $S_{entanglement}$ with finite
$\ell$ the answer will depend on the details of how $\ell$ is introduced
(sometimes called, rather misleadingly, ``scheme dependence'').  In
itself, this ambiguity might not be a problem, but for the fact that it
tends to call into question the covariance of $S$.
For example in computing entanglement entropy for a Schwarzschild black
hole, we might think to neglect all contributions from within a distance
$\ell$ of the horizon, but what does this mean?  Do we neglect modes
within a shell $\Delta{r}=\ell$ where $r$ is the Schwarzschild radial
coordinate?  Or should we be using instead of $r$, proper distance along
the spacelike surface $\Sigma$ whose entropy we want (proper distance to
a lightlike surface being not in itself well-defined)? But then the answer
will depend on how we extend $\Sigma$ away from the horizon $H$.
If for example we define $\Sigma$ as a surface of constant
Schwarzschild-time $t$ then it will go through the bifurcation
$2$-sphere.  But realistic black holes don't have such bifurcation
surfaces; and even if they did, our attention when we came to consider
the crucial question of entropy {\it increase} would have to focus on
portions of the horizon which were farther to the future.


\vbox{
   \bigskip

  \includegraphics[scale=0.5]{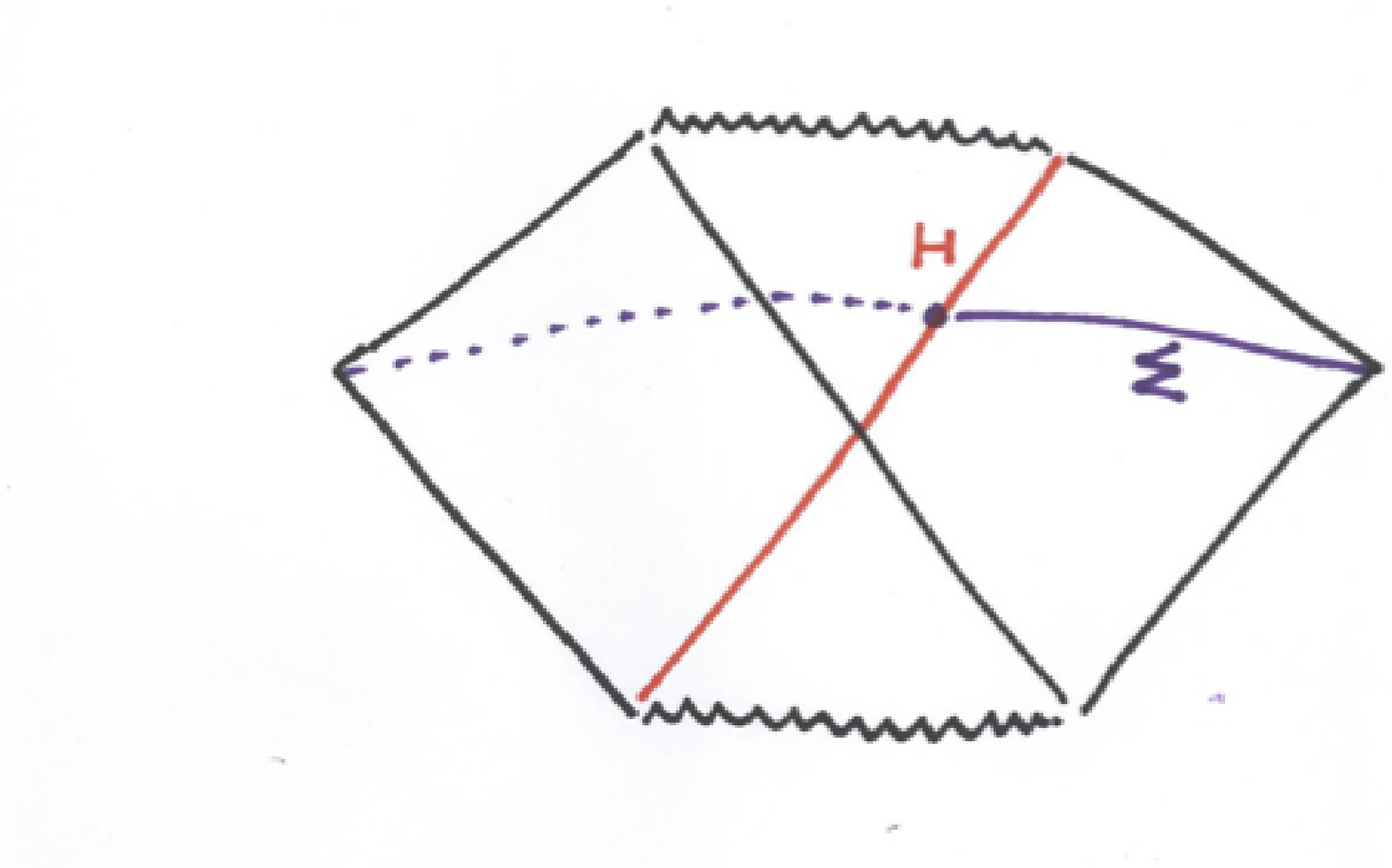}

  \Caption{{\it Figure 1.} The black hole horizon $H$ and a hypersurface $\Sigma$ whose entropy is
  desired. The dotted line extends $\Sigma$ to a Cauchy surface of the
  full spacetime. }}

\subsection{entanglement entropy in a causal set}
The difficulties we have just reviewed make it hard to define an entropy
of entanglement that is at the same time covariant, but if one replaces
spacetime by a causal set the difficulties disappear because no cutoff
is needed.  That, of course, is one reason for thinking of the causal
set as more fundamental than the continuum.  
[2] [3] [4]
But even if one is seeking
no more than a ``phenomenological cutoff'' to render the entropy finite,
a causet sprinkled into the spacetime in question offers (uniquely as
far as I know) a covariant way to obtain one.  For this reason alone, it
is worth asking whether a natural definition of entanglement entropy can
be found within causal set theory.

Let us consider then, the problem of defining a horizon-entropy in a
fixed, background causal set.
First of all one must attach a meaning to the words ``event horizon''.
Although this might seem to pose a problem because a causet contains no
subset that could play the role of a null surface, the distinction
between the interior and exterior of a black hole still makes sense 
because it depends solely on the causal order.
Since quantum entanglement concerns only the interior and exterior regions,
and not the horizon per se, there is in fact no problem.

The next step, if we were to mimic the continuum story perfectly, would
be to define the concept of a spacelike (or achronal) surface $\Sigma$~,
after which we would try to associate with each suitable $\Sigma$ some
sort of ``momentary state'' or density matrix $\rho_\Sigma$.  In fact,
there does exist a natural choice for $\Sigma$~, since we can define it as
an antichain, or better a maximal antichain (or ``slice'').
That which is not readily available in a causet, however, is anything
like $\rho_\Sigma$~: a causet-based field theory admits no obvious
notion of ``state on a hypersurface''.

More generally, none of the surface-quantities which one routinely
defines in the continuum seems to carry over naturally to an antichain,
except possibly a measure of (spatial) volume.  In fact, it seems to
be a rule of thumb that 
``purely spatial'' concepts do not live happily in a causet\footnote{$^\dagger$}
{More generally, there is evidence for the rule that it is the
 continuum concepts which are ``$C^0$-stable'' that have simple causet
 counterparts.}
including for example the concepts of induced metric $g_{jk}$ and  extrinsic
curvature $K_{jk}\sim\dot{g}_{jk}$.  For this reason a causet is not
congenial to canonical quantum gravity, albeit one {\it can} define
many surface-quantities in an {approximate} sense, including
both metrical ones and topological ones like homology groups
[5].

The question then is whether we can free $S_{entanglement}$ from
reference to a density matrix localized to a hypersurface.  In seeking
the answer, I will limit myself to what is perhaps the simplest case,
that of a free scalar field $\fihat(x)$~.
%
%
For this case a corresponding theory exists in the causet, and it has
been given both algebraic and path-integral (or ``quantum measure'')
formulations.  Inasmuch as 
the
path-integral formulation extends to the
interacting case, it is arguably preferable to the operator formulation.
Nevertheless, I will work in the latter context, first of all because
more tools are available there, and more importantly because we seem to
lack any definition of entropy couched in the language of
histories.\footnote{$^\flat$}
{I'm ignoring here the trick that computes the thermodynamic partition
 function via a Wick-rotated path integral.  That procedure is indeed
 histories-based, but the histories in question develop in imaginary
 time, and more importantly, the entropy defined thereby is limited to
 states of thermal equilibrium.}

Could it be that this lack betokens 
an inherent dependence of entropy on some notion
of ``state localized in time''?
Were that true, the search for a
histories-based definition of entropy would necessarily be futile, but
such pessimism is called into question by the global definition of
entropy we will arrive at below in the algebraic context.  If indeed a
more (Lorentzian and non-equilibrium) histories-based concept of entropy
is out there somewhere, its discovery would, to my mind, mark important
progress in the path-integral formulation of quantum theories.

\subsection{Entropy from the spacetime correlator}
Let us recall some possible definitions of entropy in the context of
quantum field theory.
Conceived algebraically (as it commonly is), a quantum field is a
collection of  
operators\footnote{$^\star$}
{In the continuum the symbol $\phi(x)$ must of course be interpreted
 formally since the field $\phi$ is at best an operator-valued
 distribution.  In the following lines I will ignore this technicality
 and whatever difficulties flow from it.  They are peculiar to
 continuous spacetime and do not seem to be germane to the causal set
 context of primary interest in this paper, within which the symbol
 $\phi^j$ belonging to element $j$ of the causet will literally denote an
 operator, albeit an unbounded one.}
$\phi(x)$ acting irreducibly in a hilbert space $\Hilb$~, 
together with
a ``global state'' represented by another operator $\rho$
(called ``the density-matrix'') such that 
$\tr\rho{A}$ 
yields the
``expectation value'' 
$\xptn{A}$
of $A$~.  One can then define
the {\it entropy} as $S=S(\rho)=\tr\rho\log\rho^{-1}$.  
(In practice the $\rho$  that enters this definition will seldom be the
``exact microscopic density matrix''.  Rather it will be some
coarse-grained version thereof whose entropy is nonzero even when the
underlying ``microscopic state'' is pure.)

Now let $R$ be an arbitrary spacetime region.  The operators $\phi(x)$
for $x\in{}R$ generate 
a subalgebra $\A_R$ of $\A$, and the expectation-value for $\A$
restricts to a similar functional on $\A_R$~.
In general, the subalgebra $\A_R$ will no longer act irreducibly in
$\Hilb$.  Assuming however, that we can represent $\A_R$
in some other Hilbert space $\Hilb_R$ in which
it does act irreducibly, and 
assuming further that we can find in $\Hilb_R$ a density matrix
$\rho_R$ such that $\xptn{A}=\tr\rho_R{}A$ for operators $A\in\A_R$,
we can go on to define $S(R)=-\tr\rho_R\log\rho_R$, 
{\it the entropy of $\rho$ relative to the region $R$}.

\REMARK  
   Let $\A$ be the algebra generated by the
 $\phi(x)$.
   As such, it is a concrete algebra of linear operators in $\Hilb$, but
   in what might be called the ``strictly algebraic'' formulations of
   quantum field theory, $\A$ is regarded as an abstract
   $\star$-algebra, while a density-matrix $\rho$ is regarded simply as
   an expectation-value-functional on $\A$.  Since the set of all such
   functionals is convex, any given $\rho$ will be 
   (either exactly or to a close approximation)
   a convex combination $\sum_\alfa p_\alfa \rho_\alfa$ 
   of extremal (or ``pure'') states $\rho_\alfa$~.
   In this abstract setting, $S(\rho)$ could perhaps be defined as 
   the infimum\footnote{$^\dagger$}
   {For $\dim\Hilb<\infty$ this is proven in [6].}
   over all such sums of the quantity 
   $\sum_\alfa p_\alfa \log p_\alfa^{-1}$~.  In this way,
   one could imagine defining the entropy $S(R)$ directly in terms
   of the expectation-value functional, without recourse to either
   $\Hilb_R$ or $\rho_R$~.

The definition we have just given 
of the entropy of an arbitrary
spacetime region $R$
will be the basis of the causal set construction we
are seeking.  First though, we need to relate the entropy of a region
to the entropy of a hypersurface.  To define the latter itself, we can
imagine repeating the same steps that led to $S(R)$, only with $R$
shrunken down to a hypersurface $\Sigma$~, 
and with $\A_\Sigma$ then being the
algebra generated (formally) by the ``initial data'' operators 
$\phi(x)$ and $\dot\phi(x)$ for $x\in\Sigma$~.


The point now is that when $\Sigma$ is a Cauchy surface for the region $R$~,
%
%
we have $\A_\Sigma=\A_R$~, 
and therefore $S(\Sigma)$ will coincide with 
the more globally defined entropy $S(R)$.  
By taking for $R$ the
so-called {\it domain of dependence} of $\Sigma$~, 
we can thus express the entropy of any desired hypersurface 
as the entropy of a spacetime region.  And this
in turn will let us express it directly in terms of the
correlation function $\xptn{\phi(x)\phi(y)}$~.

In application to a black hole spacetime, $\Sigma$ would be the exterior
portion of the hypersurface for which we desired the entanglement
entropy, and the region $R$ would then be given by
$$
    R = \hbox{future}(\Sigma) \cap \hbox{past}(H) = D^+(\Sigma) \eqno(1)
$$

\REMARK  Time-reversing the above, we could also have taken $R=D^{-}(\Sigma)$, but 
the analog of (1) would not hold with that choice, making it
apparently less convenient when transposed to a causet.


\vbox{
   \bigskip

  \includegraphics[scale=0.2]{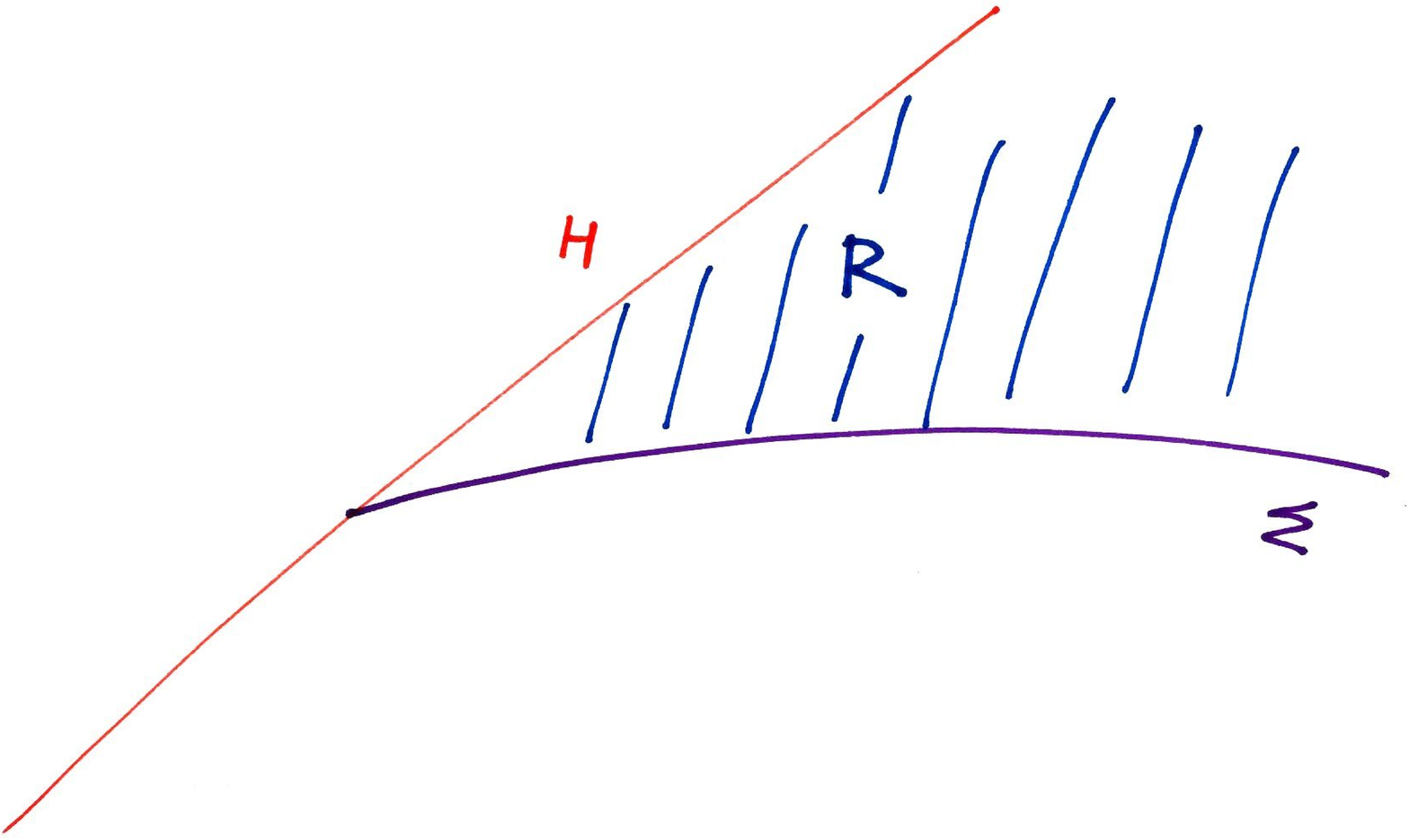}

  \Caption{{\it Figure 2.}  The entropy of the surface $\Sigma$ can be identified
  with the entropy of the region $R$}}

In summary,
{\it entanglement entropy is a special case of the entropy $S(R)$ of a spacetime region.}
Our task now is to find a simple formula for $S(R)$~, assuming, as
always, that $\phi$ is free and ``gaussian''.

\section{Sketch of a derivation of a formula}
What follows is in parts still a ``work in progress''; not every loose
end has been tied up.

We have arrived at an algebra $\A=\A(R)$ and an expectation-functional or
``state'' $\xptn{\cdot}$ thereon.  
In order to derive from these structures an entropy $S(R)$,
we will seek to represent $\A(R)$
irreducibly in a hilbert space $\Hilb=\Hilb(R)$ and to find therein a
density-matrix $\rho:\Hilb\to\Hilb$ such that $\xptn{A}=\tr \rho A$ for
all $A\in\A$.  
In what follows I will omit the reference to $R$, as if $R$ were the
full spacetime or causet as the case may be.   There is no loss
of generality in doing so, since we will not need to refer again to
spacetime points (respectively causet elements) outside of $R$. 
Also I will write ``$\phi^j$''
for an individual field value.  This notation is more convenient than
``$\phi(x)$'' and it is also more apt for a causet $C$~, where the index
$j$ runs literally from $1$ to $N$~, $N=|C|$ being the cardinality of
$C$.

\REMARK  
The condition that $\A$ act irreducibly in $\Hilb$ is crucial here.  
If we dropped it, we could, by ``purification'', always find a
representation in which $\tr\rho\log\rho$ would vanish because $\rho$
would be the projector onto a single vector in $\Hilb$.  
This would be the case in the so-called GNS representation,
and for that matter, in the representation of the $\phi^j$ in the
original, global hilbert space, assuming the original global state to
have been pure.
In view of these observations, it's somewhat disconcerting that,
strictly speaking, the existence of an irreducible representation is not
always guaranteed.  Consider for example the ``region'' of a causet
consisting of a single element $e$.  The corresponding subalgebra $\A$
will be generated by a single field-value $\phi(e)=:q$.
Now by assumption, $\xptn{q}=0\,$, 
but in general $\xptn{qq}$ will not vanish.  
If it does not, 
then $q$ cannot act irreducibly in any representation 
(because then 
  Schur's lemma $\implies$ $q=c\in\Reals$ 
  $\implies$ $q=\xptn{q}=0$
  $\implies$ $\xptn{qq}=\xptn{0}=0$).
In such a case, the entropy could not be defined by the means we have
adopted.
However, the ``strictly algebraic'' definition we remarked on earlier 
would still seem to apply and to yield $S=\infty$.
In the following, we will simply assume that an irreducible
representation does exist.  The formula (9) which we will derive
will not actually depend on this assumption if interpreted as in
(8).  It would however yield $S=0$ in our one-element
example, rather than $S=\infty$.

Because we are dealing with a free field, the commutator
of $\phi^j$ with $\phi^k$ is a $c$-number and we can write
$$
  [\phi^j,\phi^k] = i \Delta^{jk}      \eqno(2)
$$
for some real, skew matrix $\Delta^{jk}$.  We define similarly the 
{\it Wightman function} as the hermitian matrix 
$$
      W^{jk} \ideq \xptn{\phi^j \phi^k}  \ .   
$$ 
Notice that $\Delta$ is twice the imaginary part of $W$~.

Assume now that our theory is ``Gaussian'' (or ``Wickian'') in the sense
that 
the bosonic form of
{\it Wick's rule} holds with {$\xptn{\phi}=0$}:
$$
    \xptn{\phi \phi \cdots \phi} = \sum \xptn{\phi \phi}\cdots \xptn{\phi \phi}
$$
(The sum is taken over all ways of pairing the $\phi$'s, 
and within each pairing the original odering must be preserved.)
Thanks to this assumption the ``two-point function'' $W^{jk}$ determines the
theory fully.

For example its imaginary part determines 
the  ``equations of motion'' for $\phi$.
Thus 
(thinking for a moment of the theory on the entire spacetime or causet),
if the sequence $\alfa_j$ is in the kernel of $\Delta$ in the
sense that 
  $\sum \Delta^{kj}\alfa_j=0$ 
then
$A=\sum \alfa_j \phi^j$ vanishes as well.
(proof: When $\alfa\in\ker\Delta$, $A$ commutes with every $\phi^j$, as
follows from the calculation
$[\phi^j,A]=[\phi^j,\alfa_k\phi^k]=i\Delta^{jk}\alfa_k=0$~.  
But since the $\phi^j$ act irreducibly in $\Hilb$, 
this implies that $A=c{\bf{1}}$,
and then $c=\xptn{A}=\alfa_j\xptn{\phi^j}=0$ whence $A=0$~.)
This last fact yields a set of linear relations among the operators
$\phi$.  In the continuum they are just the equations of motion for
$\phi(x)$ (in our case the Klein-Gordon equation), but in the causet
they yield only relatively few conditions because the {\it exact} kernel
of $\Delta$ is typically rather small.


Now our algebra $\A$ is generated by the individual field-variables
$\phi^j$, and we can therefore characterize it fully by specifying the
relations among these generators.  But since we have taken our field to
be non-interacting, this is quite simple to do.  In addition to the
linear relations just described there are only the bilinear relations
(2) (the ``canonical commutation relations'').
Moreover, we have just seen that any zero-eigenvector of the
commutator-form $\Delta^{ij}$ results in a linear dependence among the
$\phi^j$~.  Hence, by passing to a linearly independent set of
generators, we can assume that $\Delta^{ij}$ is invertible.

It follows that there exists a basis for the $\phi^j$ of the form
$q^\alfa$, $p_\alfa$  ($\alfa=1\cdots n$)
such that 
$$
   [q^\alfa , p_\beta] = i \delta^\alfa_\beta
$$
By construction this basis block-diagonalizes $\Delta$, meaning it
block-diagonalizes the skew part of $W$~.
(Recall that $W^{jk}-W^{kj}=i\Delta^{jk}=[\phi^j,\phi^k]$~.)
I claim one can also choose it to diagonalize the symmetric part,
$$
     R^{jk} \ideq  \Re W^{jk} = \half \xptn{\braces{\phi^j,\phi^k}} \ ,
$$
hence to block-diagonalize $W^{jk}$ itself.\footnote{$^\flat$}
{In fact, diagonalization is possible in general, without assuming that
 the $\phi^j$ are linearly independent. Given any positive-semidefinite
 matrix $W$, with associated real matrices $R$ and $\Delta$ such that 
 $W=R+i\Delta/2$, one can introduce a (real) basis in which $R$ is
 diagonal and $\Delta$ is block-diagonal with $2\times2$ blocks.
 Moreover one can arrange that the non-zero matrix elements $R^{jj}=1$.
 The proof is slightly too long to reproduce here, but see [7] for a
 similar result that applies when $\Delta$ is invertible.
}
With this, our problem splits up into a product of individual problems,
each concerning a single pair ($q$, $p$) 
--- a single degree of freedom --- 
that you might imagine as position and momentum for a free particle or
harmonic oscillator
(or as the field-components $\phi^j$ for a 2-element causet, a 2-chain).
It thus suffices to evaluate the entropy in that special case.

\subsection{the entropy for a single degree of freedom}
Our task is now as follows.
Given a conjugate pair of variables $q$ and $p$ such that $[q,p]=i$~,
and given the correlators 
$$
     \xptn{qq} \quad  \xptn{pp} \quad  \Re\xptn{qp}
$$
for a Gaussian density-matrix $\rho\,$,
to find $S(\rho)=\tr \rho \log \rho^{-1}$~.
That $\rho$ is Gaussian means that in a $q$-basis it takes the form
$$
   \rho(q,q') \ideq \xptn{q|\rho|q'} 
    = 
    (cst) \exp \left(- {A \over 2}(q^2+q'^2) + {iB\over 2}  (q^2-q'^2) - {C\over 2} (q-q')^2\right)
\eqno(3)
$$
Then $S(\rho)$ must be some function of the parameters $A$, $B$, $C$,
and our task is to determine this function.

To that end, notice first that $S$ must be dimensionless and invariant
under unitary transformations.
From this it can be shown 
that $S$ can depend on the correlators only in
the combination
$ \xptn{qq}\xptn{pp} - (\Re \xptn{qp})^2 =  \det R / \det \Delta$~, 
where in this simple $2\times 2$ situation, $\Delta$ and $R$ reduce to
$$
     \Delta = 2
     \Imag   \pmatrix {\xptn{qq} & \xptn{qp} \cr
                       \xptn{pq} & \xptn{pp} \cr} 
       = 
     \pmatrix{ 0 & 1 \cr 
              -1 & 0 \cr }
$$
and
$$
     R = 
     \Re   \pmatrix { \xptn{qq} & \xptn{qp} \cr \xptn{pq} & \xptn{pp} \cr} = 
           \pmatrix { \xptn{qq} & \Re \xptn{qp} \cr \Re \xptn{qp} & \xptn{pp} \cr} 
$$
It turns out that (as a short calculation will confirm)
$$
                  \xptn{qq} \xptn{pp} - (\Re\xptn{qp})^2  = C/2A + 1/4
$$
whence the entropy depends only on the ratio $C/A$, while $B$ drops out
entirely.  But with $B$ set to 0, we can take over from
[8] the entropy calculation there, with the result\footnote{$^\star$}
{Might there be a more direct route from (3) to $S(\rho)$,
 perhaps via a diagonalization of $\rho$ or via a use of the ``replica trick''?}
$$
         - S = {{\mu\log\mu+(1-\mu)\log(1-\mu)}\over{1-\mu}} \ ,
$$
where
$$
         \mu = {{\sqrt{1+2C/A}-1}\over{\sqrt{1+2C/A}+1}}
$$
Putting the pieces together yields now
$$
    S = (\sigma+1/2) \log (\sigma+1/2) -  (\sigma-{1}/{2})\log(\sigma-{1}/{2})
  \eqno(4)
$$
where I have defined the eigenvalues of $\Delta^{-1}R$ 
(which are purely imaginary)
to be 
$\pm i\sigma$~: 
~$\hbox{spectrum}(\Delta^{-1}R)=\pm i\sigma$~.~ 

This nicely symmetrical expression is already quite simple, but 
we can simplify it still further
by swapping the eigenvalues of
$\Delta^{-1}R$ 
for those of
$\Delta^{-1}W=\Delta^{-1}R+i/2$~.~
(Recall that $W=R+i\Delta/2$.) 
Writing these latter eigenvalues 
as 
$\pm i\omega_\pm = i(\half \pm \sigma)$~,
~brings $S$ finally to the convenient form
$$
    S = \omega_{+} \log \omega_{+} - \omega_{-} \log \omega_{-}  \eqno(5)
$$
Notice in this connection that 
positivity of the correlation matrix 
(also known as the uncertainty principle) 
implies that
$\sigma\ge 1/2\,$, 
whence $\omega_{+}$ and $\omega_{-}=\omega_{+}-1$ are both $\ge0\,$.

\subsection{the entropy in full}
The formula (5) belongs to a single degree of freedom,
corresponding to one block of our block-diagonalized matrix $W$.
 Summing (5) over all the blocks then yields 
the total entropy in the form of a sum  
over the spectrum of the full matrix 
$\Delta^{-1} W$~, which for short I will call $L$, 
or more conveniently $iL$ in oder that its eigenvalues be real:
$$
         \Delta^{-1} \, W = i L             \eqno(6)
$$
Denoting the eigenvalues of $L$ by $\lambda$, we have then
$$
   S= \sum  \, \lambda \, \log |\lambda|   \eqno(7)
$$
We have seen that the eigenvalues $\lambda$ are all real, even though $L$ is in
general neither real nor symmetric.  We have also seen that each
negative $\lambda$ is paired with a positive eigenvalue $1-\lambda$.

As derived, this last formula 
relies on the fact that the $q^\alfa$ and $p^\alfa$
are all linearly independent.  When we revert to the original matrices
$W^{jk}$ and $\Delta^{jk}$, we will encounter (after their
diagonalization) a block submatrix consisting entirely of zeroes.  
%
%
In this subspace, $\Delta$ is obviously not invertible, but neither is
there any contribution to $S$.  To adapt our formula (7)
to this circumstance, we can rewrite the eigenvalue equation for
$\Delta^{-1}W$ so
that it ignores this block of zeroes.  To wit, we define the eigenvalues
$\lambda$ as the solutions of the equation
$W^{jk}v_k=i\lambda\Delta^{jk}v_k$, 
where it is understood that $\Delta^{jk}v_k$
must not vanish 
(just as $v$ must not vanish in the more traditional eigenvalue equation 
$Lv=\lambda v$).
In other words, we define the eigenvalues of $L$
as the solutions of the equation 
$$
      W v = i\lambda \Delta v \qquad\qquad (\Delta v \not= 0)\ .    \eqno(8)
$$
With this understanding (and with the usual convention that $0\log0=0$), 
we have finally
$$
     S = \tr \, L \, \log |L| \ideq  \sum_\lambda  \, \lambda \, \log |\lambda|
   \eqno(9)
$$
%
A remarkably simple formula!

\medskip

\REMARK  
Derived under the assumption that the $\phi^j$ can be represented
irreducibly in some Hilbert space, equations (8) and
(9) give a good account of the entropy when the kernels of
$\Delta$ and $R$  (or equivalently $W$) coincide.  From the condition
$W\ge0$ that $W$ is positive semidefinite, it follows that
$\ker\Delta\supseteq\ker{R}$, but the converse inclusion is not
guaranteed in general.  When it fails, no irreducible representation
exists (as seen in an earlier remark), and our entropy is to that extent ill-defined.  
If we adopt the more general ``strictly algebraic'' definition described
earlier, the resulting entropy will diverge.  By way of illustration, consider
our earlier example of a single ``field operator'' $\phi^1=q$ with
$\xptn{q}=0$ and $\xptn{qq}=s^2$
(and with $\A$ being the algebra of polynomials in $q$).
One can think of $q$ in this case as
a classical random variable with gaussian distribution function 
$1/\sqrt{2\pi s^2}\exp(-x^2/2s^2)$, the pure states corresponding to
specific sharp values $q=x$ for $x\in\Reals$~.
Since these states are uncountably infinite in number,
our ``strictly algebraic'' entropy will be infinite,
absent some sort of short-distance cutoff or regulator.

\REMARK  
Given the commutator function $\Delta^{jk}$, the construction of 
[9] and [10]
produces a distinguished ``vacuum'' by taking the two-point
function $W$ to be the positive projection of $i\Delta$, i.e. by setting
$W=\sum_n' w_n |n\rangle\langle n|$,
where the sum is over the postive eigenvalues $w_n$ of the matrix
$i\Delta=\sum_n w_n |n\rangle\langle n|$.
Evidently, the positive eigenvalues produce solutions of (8)
with $\lambda=1$, while the negative eigenvalues produce solutions with
$\lambda=0$.  The net entropy therefore vanishes, as it must since the
vacuum in question is the ``ground state'' of a Fock-type representation,
hence a pure state.



\bigskip
\noindent
Research at Perimeter Institute for Theoretical Physics is supported in
part by the Government of Canada through NSERC and by the Province of
Ontario through MRI.
%

\ReferencesBegin                             


\ref [1] Rafael D. Sorkin, ``On the Entropy of the Vacuum Outside a Horizon'',
  in B. Bertotti, F. de Felice and A. Pascolini (eds.),
  {\it Tenth International Conference on General Relativity and Gravitation (held Padova, 4-9 July, 1983), Contributed Papers}, 
  vol. II, pp. 734-736
  (Roma, Consiglio Nazionale Delle Ricerche, 1983), \lbr
  \eprint{http://www.pitp.ca/personal/rsorkin/some.papers/31.padova.entropy.pdf}
  %
  %

\ref [2] Luca Bombelli, Joohan Lee, David Meyer and Rafael D.~Sorkin, ``Spacetime as a Causal Set'', 
  \journaldata {Phys. Rev. Lett.}{59}{521-524}{1987}

\ref [3] Sumati Surya, ``Directions in Causal Set Quantum Gravity'',
 to appear in {\it Recent Research in Quantum Gravity}, 
  edited by A. Dasgupta (Nova Science Publishers NY),
  \eprint{arXiv:1103.6272v1 [gr-qc]}

\ref [4]
Fay Dowker, ``Causal sets and the deep structure of Spacetime'', 
 in
 {\it 100 Years of Relativity - Space-time Structure: Einstein and Beyond}"
 ed Abhay Ashtekar 
 (World Scientific 2005)
 \eprint{gr-qc/0508109}

\ref [5] Sumati Surya, ``Causal set topology'',
 \journaldata{Theor. Comput. Sci.}{405}{188}{2008}

\ref [6] A. Uhlmann, ``On the Shannon Entropy and Related Functionals on Convex Sets''   
 \journaldata{Reports on Math. Phys.}{1 (no. 2)}{147-159}{1970}
 %

\ref [7] Abhishek Dhar, Keiji Saito, Peter Hanggi, ``Nonequilibrium density matrix description of steady state quantum transport''
 arXiv:1106.3207
 %

\ref [8]  Luca Bombelli, Rabinder K.~Koul, Joohan Lee and Rafael D.~Sorkin, 
``A Quantum Source of Entropy for Black Holes'', 
  \journaldata{Phys. Rev.~D}{34}{373-383}{1986},

\ref [9] Steven Johnston, ``Feynman Propagator for a Free Scalar Field on a Causal Set''
 \journaldata{Phys. Rev. Lett}{103}{180401}{2009}
 \eprint{arXiv:0909.0944 [hep-th]}

\ref [10] Rafael D.~Sorkin, ``Scalar Field Theory on a Causal Set in Histories form''
 \journaldata{Journal of Physics: Conf. Ser.}{306}{012017}{2011} 
 \arxiv{1107.0698},
 \eprint{http://www.pitp.ca/personal/rsorkin/some.papers/142.causet.dcf.pdf}


\end                                        


(prog1 'now-outlining
  (Outline* 
     "\f"                   1
      "
      "
      "
      "
      "\\Abstract"          1
      "\\section"           1
      "\\subsection"        2
      "\\appendix"          1       ; still needed?
      "\\ReferencesBegin"   1
      "
      "\\ref "              2
      "\\end